\documentclass[a4paper]{jpconf}
\usepackage{graphicx}
\usepackage{amsmath}

\newcommand{\ba}{\begin{eqnarray}}
\newcommand{\ea}{\end{eqnarray}}
\newcommand{\bsub}{\begin{subequations}}
\newcommand{\esub}{\end{subequations}}
\def\b0{\beta_0}
\begin{document}
	
\title{Aspects of Shape Coexistence in the Geometric Collective Model of 
Nuclei}

\author{P. E. Georgoudis and A. Leviatan}

\address{Racah Institute of Physics, The Hebrew University, 91904, 
Jerusalem, Israel}

\ead{panos@phys.huji.ac.il, ami@phys.huji.ac.il}

\begin{abstract}
We examine the coexistence of spherical and $\gamma$-unstable deformed 
nuclear shapes, described by an SO(5)-invariant Bohr Hamiltonian, 
along the critical-line. 
Calculations are performed in the Algebraic Collective Model by introducing 
two separate bases, optimized to accommodate simultaneously 
different forms of dynamics. We demonstrate the need to modify the 
$\beta$-dependence of the moments of inertia, in order to obtain an 
adequate description of such shape-coexistence.
\end{abstract}

\section{Introduction}
The Geometric Collective Model (GCM)~\cite{GCM} 
plays a central role in the study of 
nuclear shapes and, more recently, in providing a convenient framework for 
incorporating beyond mean-field effects, essential for understanding phase 
transitions between such shapes. Of particular current experimental and 
theoretical interest, are first-order transitions involving coexistence of 
distinct shapes in the same nucleus. 
In the present contribution, we examine such coexistence of spherical and
$\gamma$-unstable deformed shapes in the vicinity of the critical-line. 
In this case, the relevant Bohr Hamiltonian involves 
a $\gamma$-independent potential, supporting two degenerate minima.
We employ an algebraic formulation of the GCM, 
the Algebraic Collective Model (ACM)~\cite{ACM05,ACM}, 
which makes exact 
numerical calculations feasible, without recourse to approximations 
such as $\beta$-rigidity and adiabaticty, hence can reveal 
general capabilities and limitations of the GCM. 

\section{The Geometric Collective Model with SO(5) symmetry}
The Bohr Hamiltonian in the quadrupole variables reads
\ba
\hat{H} = -\frac{\hbar^{2}}{2B} \left (\frac{1}{\beta^4} 
\frac{\partial}{\partial\beta} \beta^4 \frac{\partial}{ \partial \beta} 
-\frac{\hat{\Lambda}^{2}}{\beta^2}\right) 
+ V(\beta,\gamma) ~.
\label{Bohr}
\ea
Here $B$ is a mass parameter and $\hat{\Lambda}^2=-\frac{1}{\sin 3\gamma}
\frac{\partial}{\partial\gamma}\sin 3\gamma
\frac{\partial}{\partial\gamma}
+\frac{1}{4}\sum_{k}\frac{L^2_k}{\sin^2(\gamma-\frac{2}{3}\pi k)}$
is the Casimir operator of SO(5), acting on the $\gamma$ and three 
Euler angles $\Omega$. When the potential depends only on 
$\beta$, $V(\beta,\gamma)\mapsto V(\beta)$, 
the Hamiltonian has SO(5) symmetry and the wave functions can be 
separated into two parts, 
$\Psi(\beta,\gamma,\Omega)=f(\beta){\cal Y}_{\tau n_{\Delta}LM}(\gamma,\Omega)$, 
satisfying the following equations
\bsub
\ba
\hat{\Lambda}^2{\cal Y}_{\tau n_{\Delta}LM}(\gamma,\Omega) 
&=&
\tau(\tau+3){\cal Y}_{\tau n_{\Delta}LM}(\gamma,\Omega) ~,
\label{o5wf}
\\
\left [ -\frac{1}{\beta^4} \frac{\partial}{\partial\beta} 
\beta^4 \frac{\partial}{ \partial \beta}
+\frac{\tau(\tau+3)}{\beta^2} +v(\beta) \right] f(\beta) 
&=&  
\epsilon f(\beta) ~.
\label{betawf}
\ea
\esub
Here $\epsilon = \frac{2 B}{\hbar^2}E$ and 
$v(\beta)=\frac{2 B}{\hbar^2} V(\beta)$ are the reduced energy 
and potential, respectively. 
${\cal Y}_{\tau,n_{\Delta},L,M}$ 
are $SO(5)$ basis states with good $SO(5)\supset SO(3)$ 
quantum numbers $(\tau,L)$, and $n_{\Delta}$ a multiplicity label. 
By setting $\phi(\beta)=\beta^2f(\beta)$, Eq.~(\ref{betawf}) can be cast 
in the form of a radial Schr\"odinger equation
\ba
-\frac{d^2\phi}{d\beta^2} + \left [\frac{(\tau+1)(\tau+2)}{\beta^2} 
+ v(\beta)\right ]\phi = \epsilon\, \phi ~.
\label{radEq}
\ea
with an effective $\tau$-dependent potential
\ba
v^{(\tau)}_{eff}(\beta) = \frac{(\tau+1)(\tau+2)}{\beta^2} + v(\beta) ~.
\label{veff}
\ea

The ACM provides a tractable algebraic scheme for an 
exact numerical diagonalization of the Bohr Hamiltonian (\ref{Bohr}), 
in a basis of ${\rm SU(1,1)\times SO(5)}$ product wave functions, 
$R^{\lambda}_{\nu}(a \beta){\cal Y}_{\tau n_{\Delta}LM}(\gamma,\Omega)$.
The angular part are the SO(5) spherical harmonics of Eq.~(\ref{o5wf}), 
and the radial part are SU(1,1) modified oscillator wave functions 
given by~\cite{ACM05,ACM}
\ba
R^{\lambda}_{\nu}(a \beta) = 
(-1)^{\nu}\sqrt{\frac{2\nu! a}{\Gamma(\nu+\lambda)}}(a \beta)^{\lambda-1/2}
e^{-a^{2}\beta^{2}/2}L^{(\lambda-1)}_{\nu}(a^{2}\beta^{2}) 
\qquad \nu=0,1,2,\ldots
\ea
where $L^{(\lambda-1)}_{\nu}$ is a generalized Laguerre polynomial of oder $\nu$.
Any choice of $\lambda>0$ defines an orthonormal SU(1,1) basis, 
in which matrix elements of the potential can be evaluated in closed form. 
For $\lambda=\tau+5/2$, the above set reduces to the 
spherical harmonic oscillator basis. In general, the scaling parameter $a$ 
and $\lambda$ affect the width and localization of $R^{\lambda}_{\nu}(a \beta)$, 
respectively. An optimal choice of $(a,\lambda)$, enables 
a faster convergence as a function of basis size. 
For $\gamma$-independent potentials, 
with even powers of $\beta$, this implies that converged results 
can be obtained with only a few basis states (small $\nu_{\max}$) 
in the expansion of the radial wave function 
\ba
\phi(\beta;a,\lambda) = 
\sum_{\nu=0}^{\nu_{max}} c_{\nu}^{(\tau)} R^{\lambda}_{\nu}(a\beta) ~,
\label{phi-beta}
\ea
where the expansion coefficients, $c_{\nu}^{(\tau)}$, depend on $\tau$. 
The ACM has so far been tested for potentials with a single 
minimum~\cite{ACM05,ACM}. 
In the present work, we extend this approach to accommodate potentials 
$v(\beta)$ with multiple minima.

\section{Shape coexistence in the GCM with $\gamma$-independent potentials}

The dynamics associated with $\gamma$-independent potentials 
with a single minimum, has been studied extensively. For a single minimum at 
$\beta=0$, the spectrum resembles that of a spherical vibrator, describing 
quadrupole excitations of a spherical shape. The levels are arranged in 
$n_d$-multiplets composed of states with quantum numbers 
$(n_d\!=\!0,\,\tau\!=\!0,\, L\!=\!0)$, $(n_d=\!1\!,\,\tau\!=\!1,\, L\!=\!2)$, 
$(n_d\!=\!2,\,\tau\!=\!0,\,L\!=\!0;\,\tau\!=\!2,\,L\!=\!2,4)$ 
and  $(n_d\!=\!3,\,\tau\!=\!3,\,L\!=\!0,3,4,6;\,\tau\!=\!1,\,L\!=\!2)$, 
in increasing order. For a single minimum at $\beta>0$, the spectrum 
resembles that of a $\gamma$-unstable deformed roto-vibrator, with 
$\beta$ excitations of the deformed equilibrium shape. The ground band is 
composed of $\tau$-multiplets with quantum numbers 
$(\tau=0,\, L=0)$, $(\tau=1,\, L=2)$, 
$(\tau=2,\, L=2,4)$ and $(\tau=3,\, L=0,3,4,6)$, in increasing order. 
The same pattern repeats itself in excited $\beta$-bands. The dynamics 
at the critical-point of a second-order shape-phase transition, 
where the spherical minimum evolves continuously into a deformed minimum, 
can be modeled by a flat-bottomed potential. Analytic benchmarks for these 
three limits of structure are the harmonic spherical vibrator 
model~\cite{RoweWood} 
(similar to the U(5) dynamical symmetry of the interacting boson model 
(IBM)~\cite{IBM}), the $\beta$-rigid Jean-Wilets model~\cite{RoweWood} 
(similar to the SO(6) dynamical symmetry 
of the IBM), and the E(5) critical-point model~\cite{E5} 
(an infinite square-well potential), respectively. 
Characteristic signatures 
for these solvable limits are listed in Table~1.
\begin{center}
\begin{table}[t]
\caption{\label{Tab1}Characteristic properties of yrast states for 
GCM paradigms: 
a spherical vibrator~\cite{RoweWood}, E(5) critical-point~\cite{E5} 
and $\gamma$-unstable deformed rotor~\cite{RoweWood}. 
B(E2) values are in units of 
${\rm B(E2;2^{+}_{1}\rightarrow 0^{+}_{1})=100}$. 
The E2 operator is proportional to the quadrupole 
coordinate $\alpha_{2\mu}$.}
\centering
\begin{tabular}{@{}*{7}{ccc}}
\br
Observable & Spherical vibrator [U(5)] & 
\hspace{0.2cm}E(5)\hspace{0.2cm} & 
$\gamma$-unstable deformed rotor [SO(6)]\\
\mr
${\rm E(4^{+}_{1})/E(2^{+}_{1})}$ & 2 & 2.20 & 2.5 \\
${\rm E(6^{+}_{1})/E(2^{+}_{1})}$ & 3 & 3.59 & 4.5 \\
${\rm B(E2;4^{+}_{1}\rightarrow 2^{+}_{1})}$ & 200 & 168 & 143 \\ 
${\rm B(E2;6^{+}_{1}\rightarrow 4^{+}_{1})}$ & 300 & 221 & 167\\ 
\br
\end{tabular}
\end{table}
\end{center}

In the present work, we explore the dynamics of shape coexistence 
in the GCM, with an SO(5)-invariant Bohr Hamiltonian. 
We focus the discussion to the critical-line of a first-order phase 
transition involving the coexistence of spherical and $\gamma$-unstable 
deformed shapes. This can modeled by the following sextic 
potential in the Bohr Hamiltonian
\ba
v(\beta) &=& v_0\beta^2(\beta^2-\b0^2)^2 ~,
\label{v-beta}
\ea
where $v_{0} = \frac{2 B}{\hbar^2}V_0$ is the reduced strength.
$v(\beta)$ is an even function of $\beta$ 
and is independent of $\gamma$, ensuring an SO(5) symmetry for the 
Hamiltonian. For $v_{0}>0$, it supports two degenerate global minima, 
spherical ($\beta=0$) and deformed ($\beta=\b0>0$), at zero energy. 
A local maximum at $\beta=\frac{1}{\sqrt{3}}\b0$ creates a barrier of 
height $B_h = \frac{4}{27}v_0\b0^6$, 
separating the two minima. The two wells are 
asymmetric with different stiffness, 
$v''(\beta\!=\!\b0)=4v''(\beta\!=\!0)=8v_0\b0^4$.
The barrier-height and potential stiffness are related, 
both increasing linearly with $v_0$ and as a power-law with $\b0$.
The spherical well is finite, while the deformed well has a finite 
interior boundary and an exterior wall rising as $\beta^6$ for 
$\beta\rightarrow\infty$. Both wells become wider for higher energy. 

Finite-N aspects of the dynamics in such potentials have been 
studied~\cite{LevGav17} in the framework of the IBM, 
based on a compact U(6) spectrum generating 
algebra. Here we examine the analogous dynamics in the 
GCM, which has an inherent non-compact algebraic structure. 
The sextic potential in question, does not belong to a class of 
solvable nor 
quasi-solvable potentials, hence the radial equation~(\ref{radEq}) 
necessitates a numerical solution. 
For that purpose, we employ the ACM approach.
To facilitate the identification and convergence of different types of 
states, we diagonalize the Bohr Hamiltonian with 
$v(\beta)$, Eq.~(\ref{v-beta}), 
in two different ${\rm SU(1,1)\times SO(5)}$ bases
\bsub
\ba
\Psi_{s} &=&
\beta^{-2}\phi_{s}(\beta;a_{s},\lambda_{s})
{\cal Y}_{\tau n_{\Delta}LM}(\gamma,\Omega) ~,
\label{Psi-s}
\\
\Psi_{d} &=&
\beta^{-2}\phi_{d}(\beta;a_{d},\lambda_{d})
{\cal Y}_{\tau n_{\Delta}LM}(\gamma,\Omega) ~.
\label{Psi-d}
\ea
\label{Psi-sd}
\esub
Here the radial wave functions are defined as in Eq. (\ref{phi-beta}), 
and are characterized by different parameters 
$(a_s,\lambda_s)\neq(a_d,\lambda_d)$.
In particular, the basis $\Psi_s$ of Eq.~(\ref{Psi-s}) 
[$\Psi_d$ of Eq.~(\ref{Psi-d})]
is appropriate for spherical [deformed] type of states,
associated with the spherical [deformed] minimum, and consequently,   
$\lambda_s << \lambda_d$.
The convergence is verified by examining the energies, B(E2) values 
and orthogonality of the calculated states, $\Psi_s$ and $\Psi_d$. 
\begin{figure}[t]\centering
\begin{minipage}[t]{0.47\linewidth}
\includegraphics[width=\linewidth]{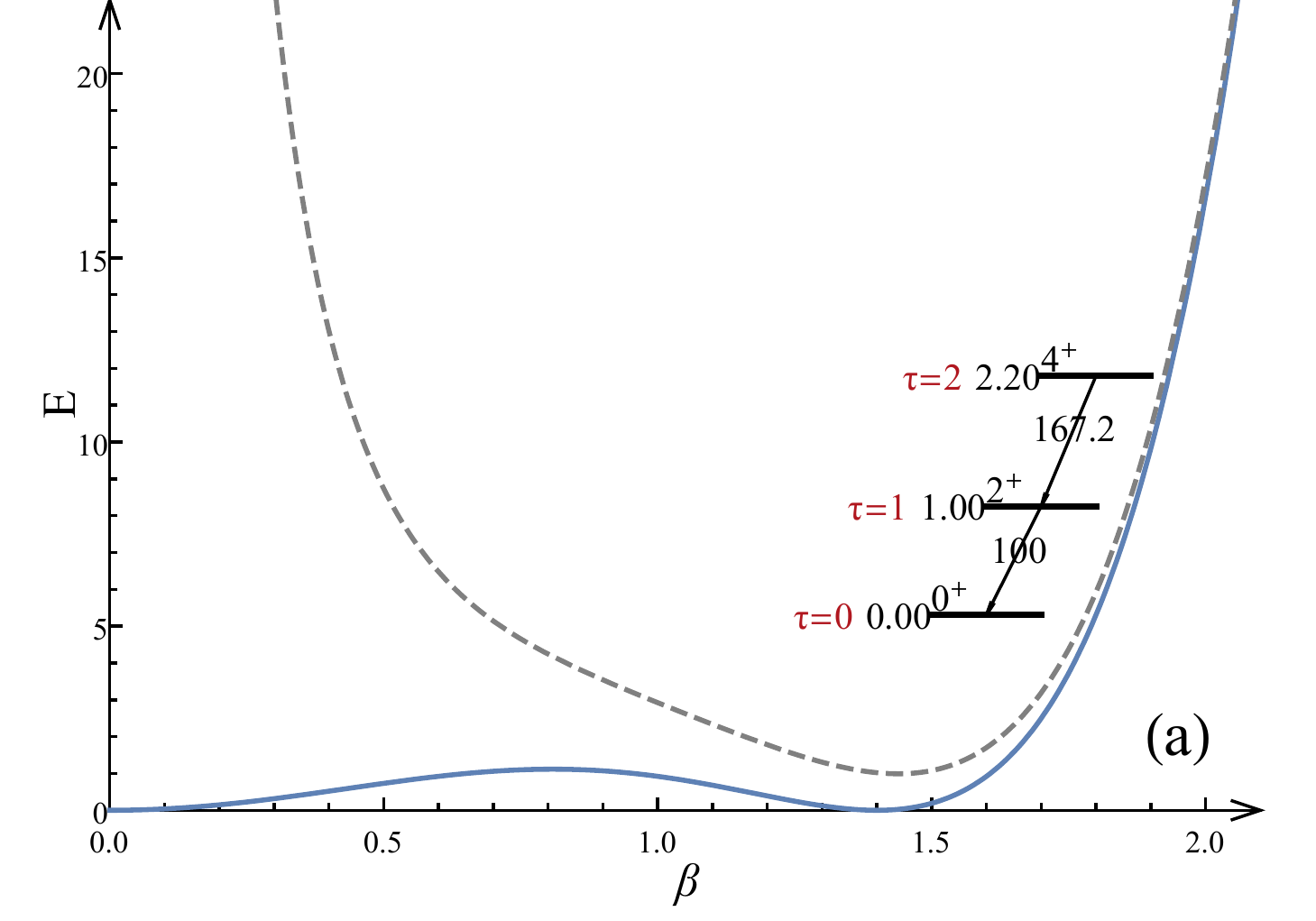}
\end{minipage}\hfill
\begin{minipage}[t]{0.47\linewidth}
\includegraphics[width=\linewidth]{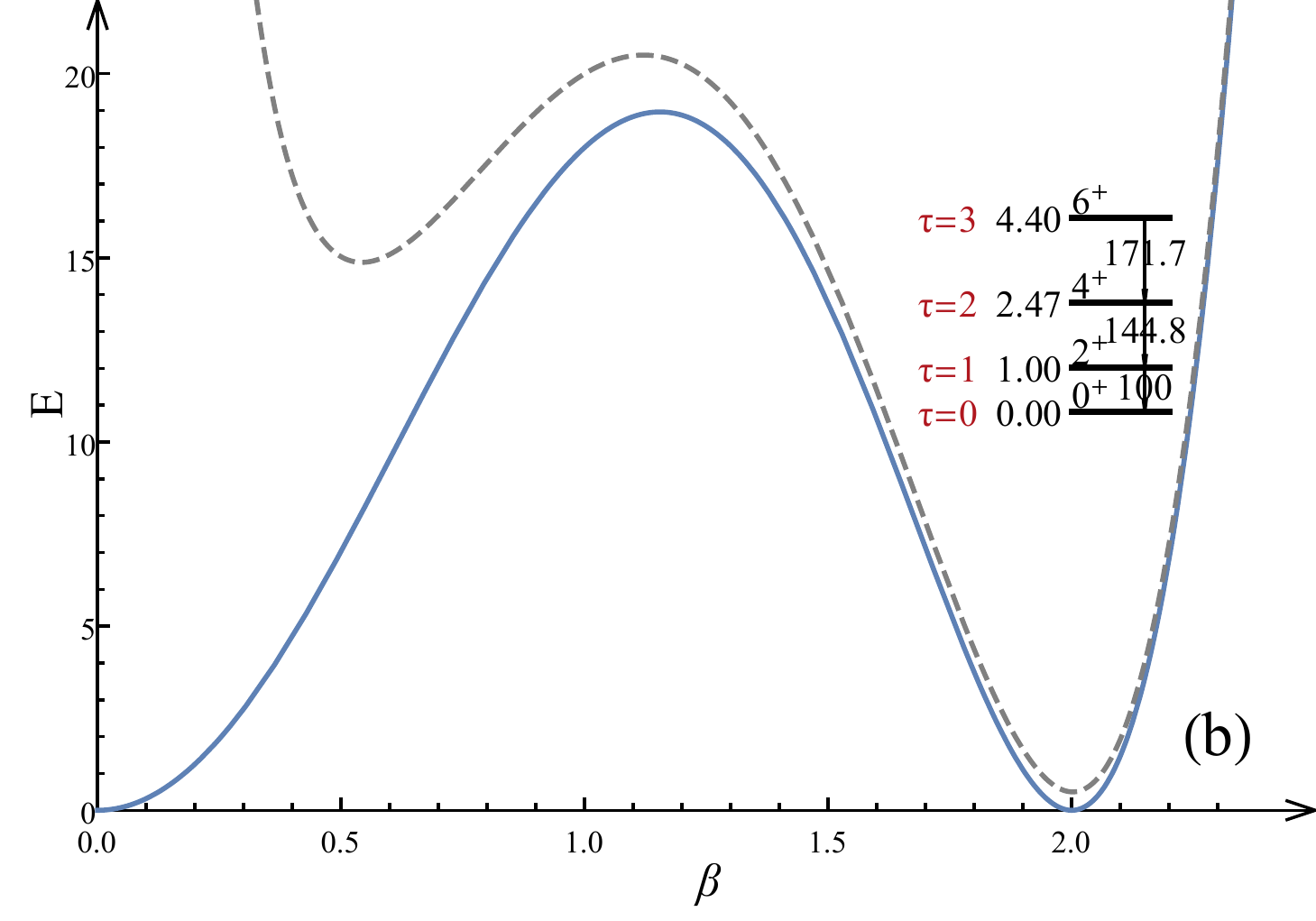}
\end{minipage}
\begin{minipage}[t]{0.47\linewidth}
\includegraphics[width=\linewidth]{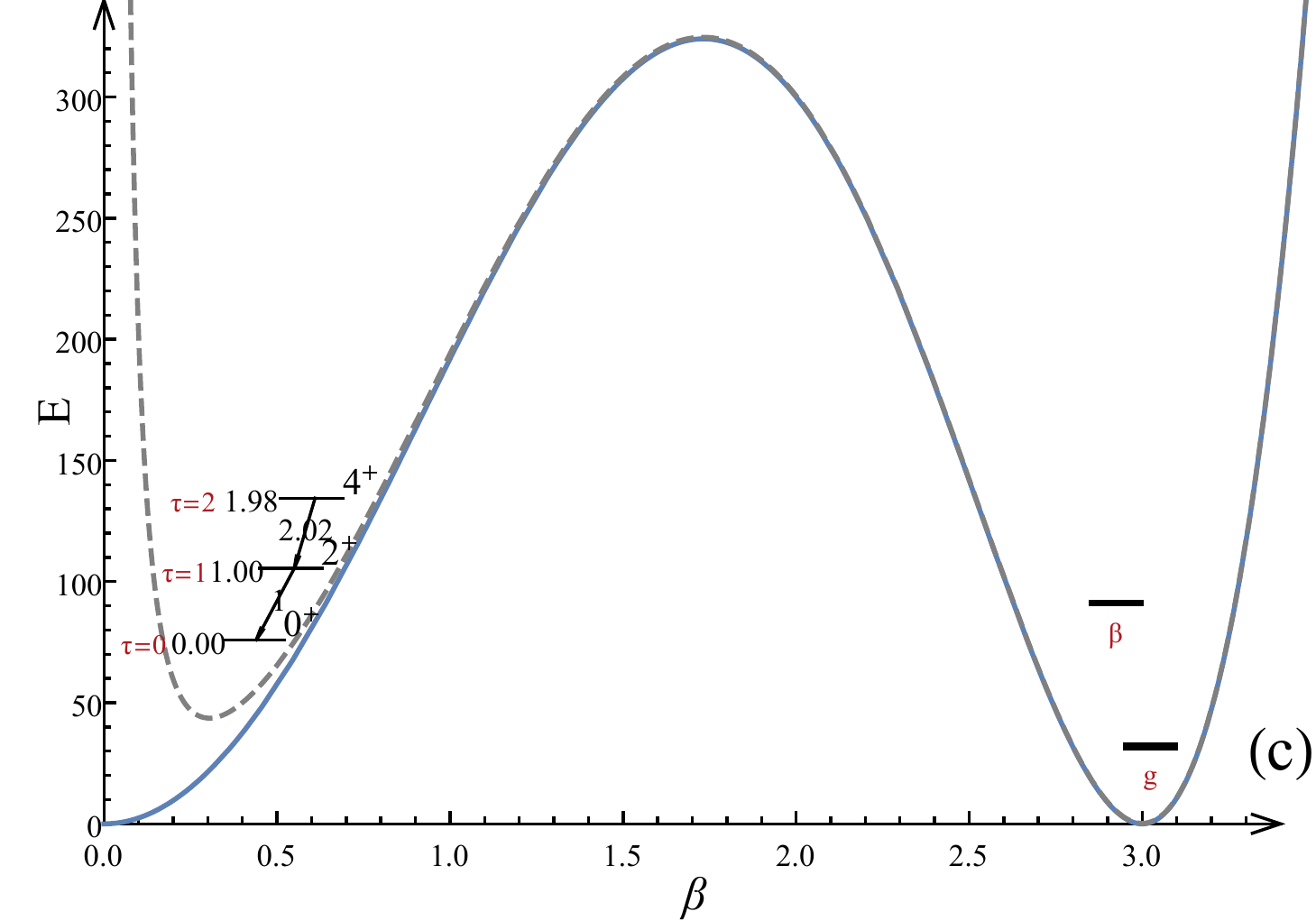}
\end{minipage}\hfill
\begin{minipage}[t]{0.47\linewidth}
\vspace{-5.1cm}
\caption{Spectrum of the Bohr Hamiltonian overlayed on the potentials
$v(\beta)$, Eq.~(\ref{v-beta}), (blue solid lines) and 
$v^{(\tau=0)}_{eff}(\beta)=v(\beta)+2/\beta^2$, Eq.~(\ref{veff}),  
(grey dashed lines), with parameters 
(a)~$(\beta_{0},v_{0})=(1,1.4)$, 
(b)~$(\beta_{0},v_{0})=(2,2)$, 
(c)~$(\beta_{0},v_{0})=(3,3)$, 
corresponding to low, intermediate and high barriers, respectively. 
Absolute energies are in units of $\frac{\hbar^2}{2B}=1$. 
Numbers to the left of the drawn levels, indicate relative energies.}
\end{minipage}
\end{figure}  

\vspace{-5pt}
\section{Evolution of structure along the critical-line}

The spectrum and eigenstates of the Bohr Hamiltonian is governed by 
the competition between the potential and kinetic-rotational terms. 
The potential term is attractive and tends to 
localize the wave functions in the vicinity of its minima. In general, 
``deepening'' a potential lowers the energies of levels confined within it, 
while ``narrowing'' a potential raises the level energies.
The kinetic term, $\frac{1}{\beta^2}\hat{\Lambda}^2$, is repulsive and tends 
to delocalize the wave-functions. Its strength depends on $\tau$ and 
the $1/\beta^2$ dependence energetically penalizes small $\beta$ values and 
tends to ``push'' the wave function towards larger $\beta$. 
This affects strongly the spherical minimum in the effective 
potential $v^{(\tau)}_{eff}(\beta)$, Eq.~(\ref{veff}), 
shifting its position to larger values of $\beta$ and distorting its shape. 
In contrast, the kinetic term has a marginal effect on the shape of 
the deformed well, yet it governs the rotational splitting of 
states in the associated ground and excited bands. 
These effects are seen clearly in Fig.~1, 
displaying the energy spectrum overlayed on the potentials $v(\beta)$ and 
$v^{(\tau=0)}_{eff}(\beta)$, for representative values of $(v_0,\b0)$. 
Fig.~1(a) corresponds to the case of a low-barrier. 
The energy levels are well above the barrier and experience essentially a 
flat bottomed potential. 
The spectrum resembles 
that of the E(5) critical-point model (see Table~1 for comparison). 
Higher levels 
show the spectral character of a $\beta^6$ potential. 
The kinetic term present in the effective potential, 
completely washes out the spherical minimum. 
Fig.~1(b) corresponds to the case of an intermediate-barrier.
Here a spherical minimum develops in $v^{(\tau=0)}_{eff}(\beta)$ 
but is shifted to higher energy and is distorted by the kinetic term 
to such an extent, that it does not support bound states. 
The spectrum consists only of deformed type of states, 
forming a ground band with and SO(6)-like structure. 
Fig.~1(c) corresponds to the case of a high barrier. 
Here both the spherical and deformed minima are well developed and 
support bound states localized in their vicinity. 
The deformed states comprise the ground and $\beta$ bands. 
Their rotational splitting, shown in Fig.~2, 
exhibits an SO(6)-like character. The spherical states 
are arranged in $n_d$-multiplets and exhibit a U(5)-like structure. 

\vspace{-5pt}
\section{Limitation of the standard GCM Hamiltonian and a possible 
resolution}

Figs. 1(c)-2, demonstrate a limitation of the standard GCM Hamiltonian, 
namely, very different energy scales for the rotational 
and vibrational excitations. 
A pronounced coexistence of many spherical and deformed states 
requires the two wells to be deep with a high-barrier in-between. 
For the potential under study, this necessitates a large deformation $\b0$. 
The states localized within the deformed potential 
have an average deformation $\langle\beta^2\rangle$ of order $\b0^2$, 
increasing with $\tau$, hence experience 
a small rotational splitting of order $\tau(\tau+3)/\b0^2$, 
with noticeable centrifugal stretching. 
For the same $\b0$, the stiffness of the deformed minimum is large, 
hence the $\beta$-bandhead energy is high, 
of order $\epsilon_{\beta}\approx 4\b0^{2}v_{0}$. 
The kinetic term shifts the position of 
the spherical minimum to a small but non-zero value of $\beta$, hence 
the lowest spherical $0^{+}_{s}$ state 
experiences a shift of order $2/\langle\beta^2\rangle$ 
to higher energy.
For the example considered in Figs.~1(c) and 2, 
$E(2^{+}_1)\!-\!E(0^{+}_1)\!=\! 0.45$, while 
$\frac{E(0^{+}_{\beta})-E(0^{+}_{1})}{E(2^{+}_{1})-E(0^{+}_{1})}\!=\!131.06$
and $\frac{E(0^{+}_{s})-E(0^{+}_{1})}{E(2^{+}_{1})-E(0^{+}_{1})}\!=\!98.90$, 
{\it i.e.}, the resulting rotational and vibrational scales 
differ by more than two-orders of magnitude. Such a difference is at variance 
with the experimentally observed patterns of coexistence in nuclei.
\begin{figure}[t]\centering
\begin{minipage}[t]{0.47\linewidth}
\includegraphics[width=\linewidth]{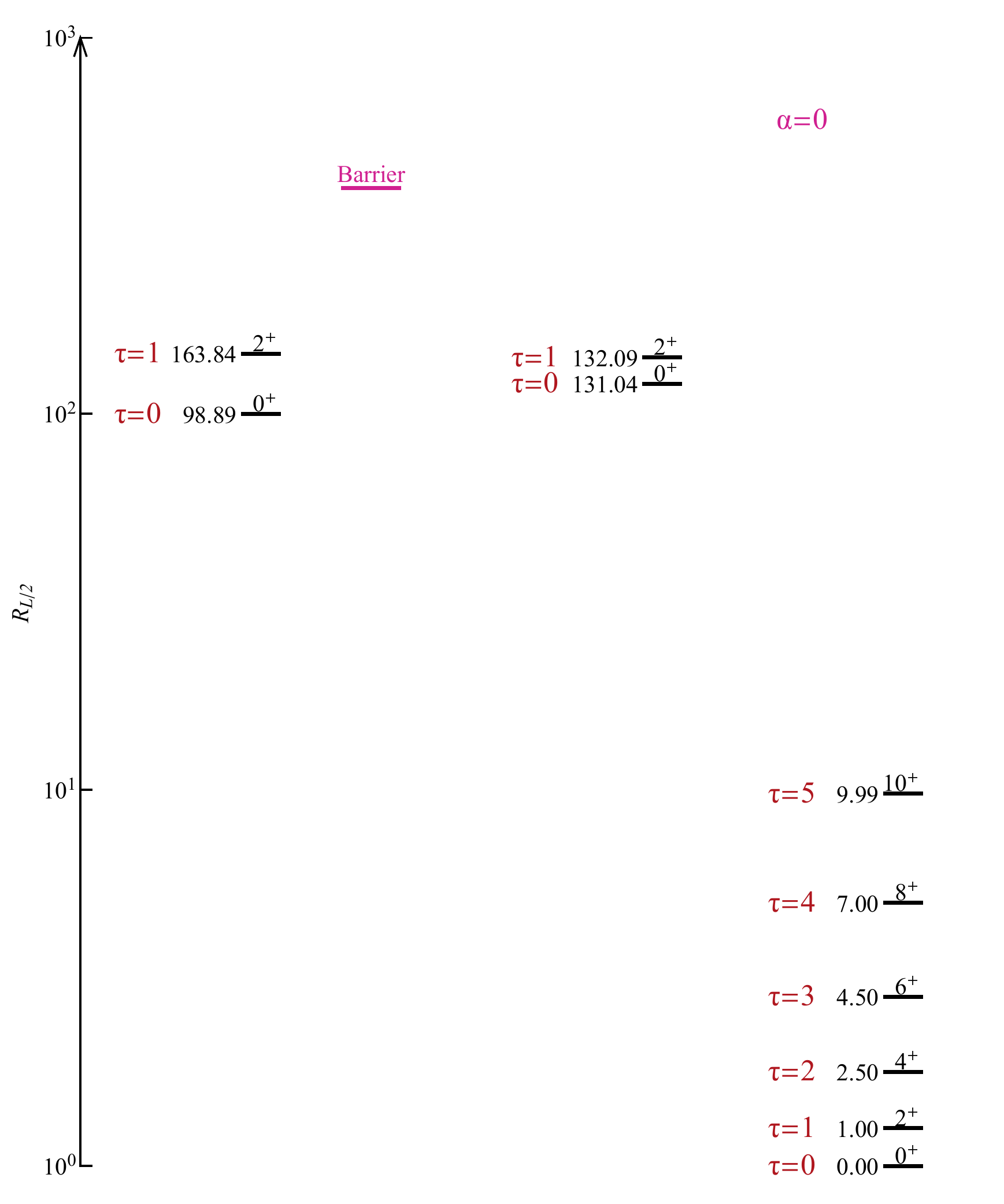}
\caption{
Selected levels in the ground band (right), 
$\beta$-band (middle) and spherical states (left) for 
the Bohr Hamiltonian with $(\b0,v_0,\alpha)\!=\!(3,3,0)$. 
Energies are in units of 
$R_{L/2}=\frac{E(L)-E(0^{+}_{1})}{E(2^{+}_{1})-E(0^{+}_{1})}$.}\label{Fig2}
\end{minipage}\hfill
\begin{minipage}[t]{0.47\linewidth}
\includegraphics[width=\linewidth]{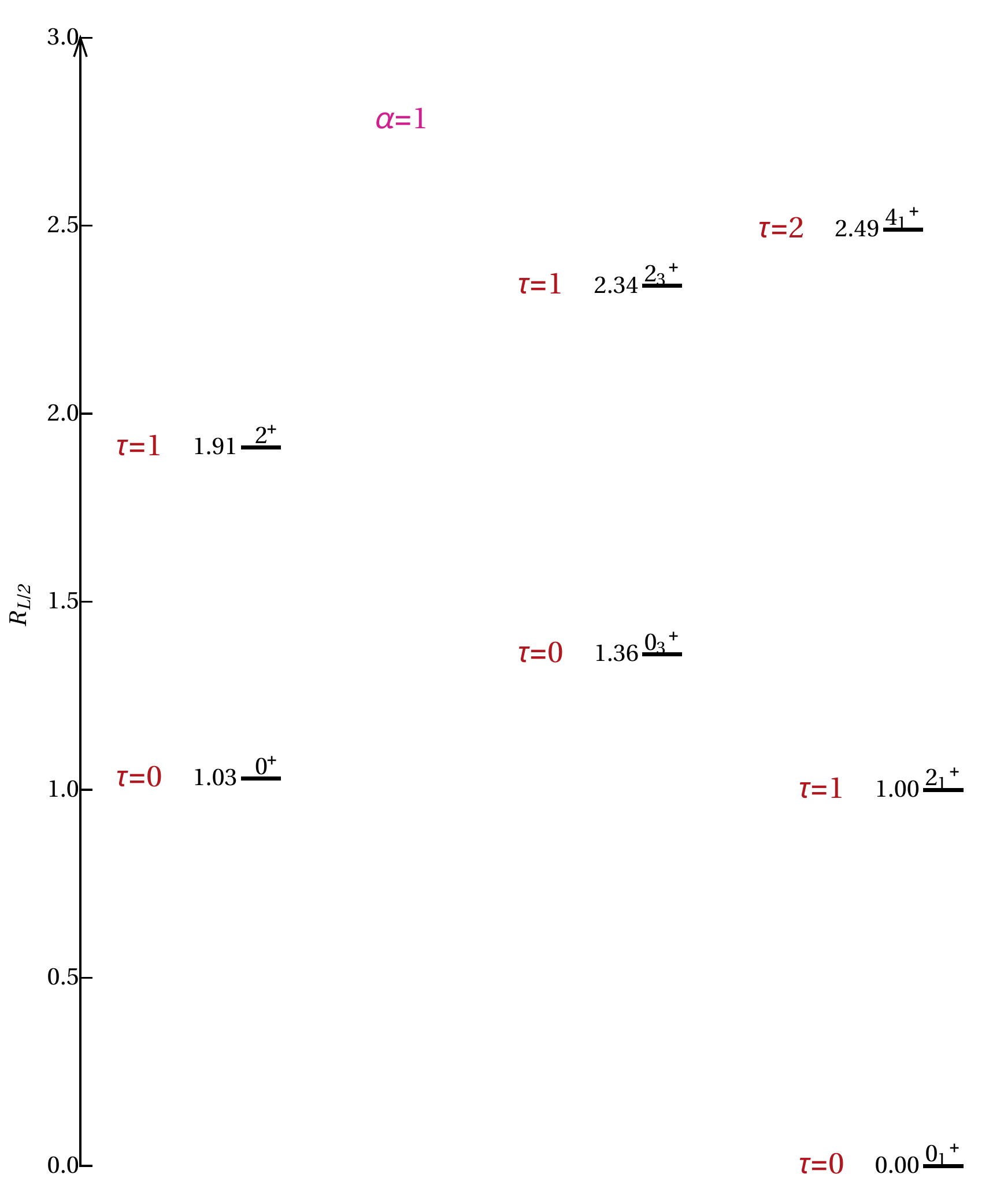}
\caption{As in Fig.~2, but for a Bohr Hamiltonian with 
$(\b0,v_0,\alpha)=(3,3,1)$ in Eqs.~(\ref{v-beta}) and (\ref{mI}).
Note the linear energy scale as opposed to the logarithmic energy scale 
in Fig.~2.}\label{Fig3}
\end{minipage}
\end{figure}  

The origin of the problem can be traced to the 
quadratic $\beta$-dependence of the irrotational moments of inertia 
in the kinetic term, $\tfrac{1}{\beta^2}\hat{\Lambda}^{2}$. 
A possible resolution is to allow a departure from such 
a $\beta^2$ behavior. 
In the present study, we consider the following substitution
\ba
\frac{1}{\beta^2}\hat{\Lambda}^{2}
\rightarrow
\frac{(1+\alpha\beta^2)^2}{\beta^2}\hat{\Lambda}^{2} ~.
\label{mI}
\ea
This particular choice is motivated by 
previous studies within the GCM~\cite{Bona11,Panos14} 
and the classical limit of the IBM~\cite{Bona15}. 
For $\alpha\neq 0$, this introduces an additional term, 
$\alpha(2 + \alpha\beta^2)\tau(\tau+3)$, in the radial equation, 
which affects only the levels with $\tau\neq 0$. 
Typical spectrum and E2 decay pattern, obtained with $\alpha=1$, are 
shown in Figs.~3-4. 
Many states now occur below the barrier, with comparable rotational 
and vibrational scales. The spherical states exhibit a U(5)-like structure, 
with strong $\Delta n_d=\pm 1$ E2 decays. The deformed states 
(comprising the ground and $\beta$-bands) maintain 
an SO(6)-like structure, with strong (weak) intra-band (inter-band) 
transitions. The E2 decays between the spherical and deformed type 
of states are extremely weak, reflecting the impact of the high barrier.
The ability of a single Hamiltonian 
to accommodate simultaneously states with different symmetry character, 
reinforces the view that partial symmetries can play a role in 
the phenomena of shape coexistence~\cite{LevGav17,Lev07,LevDek16}.
\begin{figure*}[t!]\centering
\includegraphics[scale=0.65]{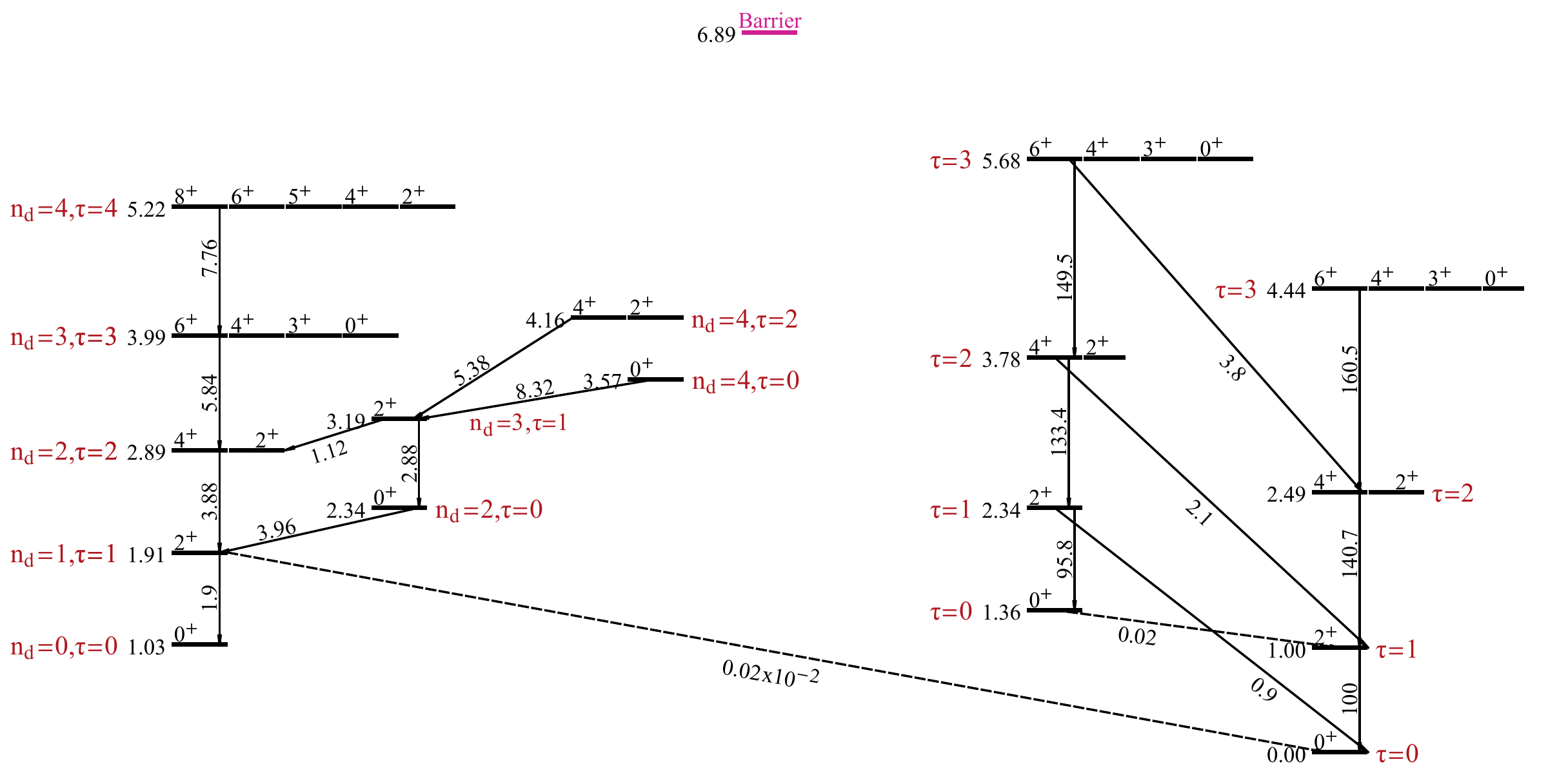}
\caption{Spectrum of the Bohr Hamiltonian with 
$(\b0,v_{0},\alpha)=(3,3,1)$ in Eqs.~(\ref{v-beta}) and (\ref{mI}). 
Converged results for spherical (left side) and deformed 
(right side) type of states, are obtained by employing 
${\rm SU(1,1)\times SO(5)}$ bases, Eq.~(\ref{Psi-sd}), 
with $(a_s,\lambda_s)=(4,2.5)$ and 
$(a_d,\lambda_d)=(4.1,140.5)$, respectively.
$L\!=\!0^{+}_{1},\,0^{+}_{3}$ are the bandhead states of the ground and 
$\beta$ bands, with $\sqrt{\langle \beta^{2} \rangle} =2.97,\,2.93$, 
respectively. $L\!=\!0^{+}_{2}$ is the spherical ground state 
with $\sqrt{\langle \beta^{2} \rangle} =0.41$. }\label{Fig4}
\end{figure*}

\ack
This work is supported by the Israel Science Foundation Grant No. 586/16. 
P.E.G. acknowledges the Golda Meir Fellowship Fund for partial support. 
We thank M.A. Caprio for assistance in using his ACM numerical code.

\end{document}